\documentstyle[12pt]{article}
\input epsf
\newcommand{\beq}{\begin{equation}}
\newcommand{\eeq}{\end{equation}}
\newcommand{\beqa}{\begin{eqnarray}}
\newcommand{\eeqa}{\end{eqnarray}}
\newcommand{\ba}{\begin{array}}
\newcommand{\ea}{\end{array}}
\newcommand{\CR}{\nonumber \\}

\newcommand{\cN}{{\cal N}}
\newcommand{\bo}[1]{\mbox{\boldmath $\bf#1$}}
\begin{document}
\begin{titlepage}
\begin{flushright}
{\tt hep-th/0105291} \\
UT-942 \\
May, 2001
\end{flushright}
\vspace{0.5cm}
\begin{center}
{\Large \bf
String Webs and Curves of Marginal Stability \\
	     in Five-Dimensional $E_N$ theories on $S^1$ \par}
\lineskip .75em
\vskip2.5cm
{\large Yukiko Ohtake}
\vskip 1.5em
{\large\it Department of Physics, University of Tokyo \\
\vspace{1mm}
Tokyo 113-0033, Japan}
\end{center}
\vskip3cm
\begin{abstract}

We study curves of marginal stability (CMS) in five-dimensional $\cN=1$ 
$E_N$ theories compactified on a circle using the D3-brane probe 
realization.
In this realization, BPS states correspond to string webs in the affine 
$E_N$ 7-brane background and junction positions of the webs determine
CMS. We find that there exist string webs involving infinitely many 
junctions. Consequently the $E_N$ theories have infinitely many CMS.
We also find that there exists a transition from open strings to string 
webs involving loops.
The transition describes a new phenomenon occurring on CMS.

\end{abstract}
\end{titlepage}
\baselineskip=0.7cm
\section{Introduction}
It is well known that BPS spectrum of four-dimensional $\cN\!=\!2$ 
gauge theories jumps non-perturbatively as the moduli of the theory 
are varied.
This phenomenon was first studied in $SU(2)$ Yang--Mills theory 
and the study has been extended to $SU(2)$ QCD\cite{SW,BF2}.
In addition the phenomenon has been investigated using stringy 
realizations of the gauge theories\cite{SR}.
Especially the D3-brane probe realization of $SU(2)$ Yang--Mills theory 
gives us a transparent understanding of the phenomenon\cite{Fy, BFy, MNS}.

In the D3-brane probe realization, $SU(2)$ Yang--Mills theory arises 
as the world-volume theory on a D3-brane in a background of an 
O7-plane\cite{Sen,BDS}.
The position of the D3-brane on $\bo{P}^1$ which is the space transverse 
to the O7-plane corresponds to the moduli parameter.
When the D3-brane is located at the O7-plane, gauge symmetry is enhanced
to $SU(2)$.
Thus the position of the O7-plane corresponds to the singularity in the 
classical moduli space.
In accordance with the singularity splitting in field theory,
the O7-plane splits into two mutually non-local 7-branes.
Hence there appear the 3-string junctions connecting these 
7-branes and the D3-brane.
These string webs correspond to dyons and W-bosons in field theory.
The position of the junction should be located on a circle passing 
through the 7-branes to balance the string tensions\cite{BFy, MNS}.
When the D3-brane is located on the circle, 3-string junctions become 
marginally stable.
When the D3-brane is located inside the circle, 3-string junctions 
are no longer stable and they decay into open strings emanating from 
each 7-brane. 
These behaviors of string webs reproduce the jump of BPS spectrum.

When we add $N_f\leq 4$ D7-branes to the O7-plane background,
$SU(2)$ gauge theory with $N_f$ fundamental matters arises 
on the D3-brane.
The positions of D7-branes correspond to masses of the matter fields.
If the masses are turned on, D7-branes separate from other 7-branes. 
Thus string webs connecting three or more 7-branes and 
the D3-brane generally involve multiple 3-string junctions. 
Consequently infinitely many curves of marginal stability 
(CMS) appear in the theories\cite{O}.

In this paper we will extend the study to five-dimensional $SU(2)$ gauge 
theory with $5\leq N_f\leq 7$ fundamental massless matters compactified 
on $S^1$.
Especially we will consider the strong coupling limit of the theory
whose global symmetry $SO(2N_f)\times U(1)$ is enhanced to 
$E_{N_f+1}$\cite{S}.
Classically the five-dimensional theory is realized on a D4-brane in the 
background of an O8-plane and $N_f$ D8-branes.
Compactifying the system on $S^1$ along the D4-brane and taking T-dual,
we see that the theory on $S^1$ appears on a D3-brane in the background 
of two O7-planes and $N_f$ D7-branes. 
As each O7-plane splits into two 7-branes, the background includes 
$(N_f+4)$ 7-branes.
$(N_f+3)$ of them collapse to realize the $E_{N_f+1}$ symmetry.
It is known that the 7-brane background realizes the affine $E_{N_f+1}$
algebra\cite{DW,DHIZ}.

We will derive the CMS of string webs in the 7-brane background. 
Since the string webs are dual to D-branes 
in IIA theory compactified on a Calabi--Yau 3-fold with a shrinking 
del Pezzo 4-cycle\cite{HI,MOY}, the CMS would add our understanding 
of D-brane stability\cite{D}.
In addition we will find that some string webs draw a loop around the 
7-branes and intersect themselves.
The vertex of strings is located on a curve of marginal stability 
similar to the junction position of a 3-string junction.
When the D3-brane is located outside the curve, the string web has
a modulus which corresponds to the size of the loop.
The loop disappears from the string web when the D3-brane is located 
inside the curve.
Thus the moduli space of a BPS state jumps as the moduli of the 
theory are varied.

The organization of this paper is as follows.
In section 2 we review the D3-brane probe realization of 
$E_N$ theories and the duality map between IIA D-branes and string webs.
In section 3 we construct 3-string junctions and determine their CMS. 
In addition we observe that some string webs intersect themselves.
In section 4 we study the behavior of the self-intersecting string webs.
Consequently we find that the moduli space of a string web changes 
as the position of the D3-brane is varied.
In section 5 we deform 3-string junctions using the Hanany--Witten 
effect. As a result we find string webs involving multiple junctions. 
Their CMS are determined similar to those of 3-string junctions. 

\section{D3-brane realizations of $E_N$ theories}

The affine $E_N$ 7-brane backgrounds are described by elliptic curves,
\beqa
E_8 &:& y^2 = x^3 + R^2 u^2 x^2 - 2 u^5,\CR 
E_7 &:& y^2 = x^3 + R^2 u^2 x^2 + 2 u^3 x, \label{eqn:ell}\\
E_6 &:& y^2 = x^3 + R^2 u^2 x^2 - 2 R i u^3 x - u^4,\nonumber
\eeqa
where $R$ is a parameter with the mass dimension $-1$ and 
$u$ is a complex coordinate of $\bo{P}^1$ which is the space transverse 
to 7-branes. 
The positions of the 7-branes on $\bo{P}^1$ are represented by zeros 
of the discriminant
\beqa
E_8 &:& \Delta (u) = -4 u^{10} ( 2 R^6 u -27 ),\CR
E_7 &:& \Delta (u) = -4 u^9 ( R^4 u -8),\\
E_6 &:& \Delta (u) = -i u^8 ( 4 R^3 u + 27 i ).\nonumber
\eeqa
The positions are $z=0$ and $z=1$ in the variable
\beq
z = \frac{2}{27} R^6 u, \,\,\,\,
    \frac{1}{8} R^4 u, \,\,\,\,
    \frac{4i}{27} R^3 u, \,\,\,\,
\eeq
for $E_8$, $E_7$, $E_6$, respectively. 
Following the convention in \cite{MOY},
we identify $(N+2)$ 7-branes located at $z=0$ with $[1,0]^N[3,-1][3,-2]$ and 
a 7-brane at $z=1$ with $[0,1]$.
The 7-branes at $z\!=\!0$ realize the $E_N$ algebra
and the 7-brane at $z=1$ is responsible for extending $E_N$ to the affine 
$E_N$ \cite{Jo,DW,DHIZ}.

At each point on the $z$-plane the curve (\ref{eqn:ell}) describes a torus.
The torus periods are given by 
\beq
\varpi(z)=\oint_A\frac{dx}{y}, \,\,\,\, 
\varpi_D(z)=\oint_B\frac{dx}{y},
\eeq
where $A$ and $B$ are the homology cycles on the torus.
These periods determine the dilaton $\phi(z)$, the RR zero-form 
field $\chi(z)$ and the metric on the $z$-plane $ds$ in the following 
way\cite{GSVY,Sen2};
\beqa
\tau(z) &=& i e^{-\phi(z)}+\chi(z)=\frac{\varpi_D(z)}{\varpi(z)},\\
ds^2 &=& \mbox{Im}\,\tau(z) \left|\varpi(z)\,dz\right|^2. \label{eqn:metric}
\eeqa

The explicit forms of $\varpi_D$ and $\varpi$ are calculated in \cite{MOY}.
For the later use we give here the results for $|z|<1$;
\beq
\left(\ba{c}
\varpi_D(z)\\\varpi(z)
\ea\right)
=\frac{\pi R^{\frac{1-\alpha}{\alpha}}}{2(2\alpha-1)}
\left(\ba{cc}
2\xi_2 & -2\xi_1\\
-\frac{\omega}{\sin\pi\alpha}\xi_2 & \frac{\bar{\omega}}{\sin\pi\alpha}\xi_1
\ea\right) 
\left(\ba{c}
z^{-(1-\alpha)}F(\alpha,\alpha ; 2\alpha ; z)\\
z^{-\alpha} F(1-\alpha, 1-\alpha ; 2(1-\alpha) ; z)
\ea\right),
\label{eqn:period}
\eeq
where $\alpha=1/6$, $1/4$, $1/3$ for $E_8$, $E_7$, $E_6$, and
\beq
\omega = e^{i\pi(\frac{1}{2}-\alpha)}, \,\,\,\,
\xi_1  = \frac{\Gamma(2\alpha)}{\Gamma^2(\alpha)},\,\,\,\,
\xi_2  = \frac{\Gamma(2-2\alpha)}{\Gamma^2(1-\alpha)}.
\eeq
$F(\alpha,\beta;\gamma;z)$ is the hypergeometric function defined by 
\beq
F(\alpha,\beta;\gamma;z)=
\frac{\Gamma(\gamma)}{\Gamma(\alpha)\Gamma(\beta)}
\sum_{n=0}^{\infty}
\frac{\Gamma(\alpha+n)\Gamma(\beta+n)}{\Gamma(\gamma+n)}\frac{z^n}{n!}.
\eeq
The branch cut of the function is taken to be along the real axis of the 
$z$-plane to $\infty$.

Now we introduce a D3-brane probe parallel to the 7-branes.
The world volume theory on the D3-brane is the five-dimensional $E_N$ 
theory compactified on $S^1$\cite{YY}.
The radius of $S^1$ is identified with $R$ in (\ref{eqn:ell}).
The moduli parameter corresponds to the position of the D3-brane on 
the $z$-plane.
$U(1)$ gauge symmetry is apparently realized on the D3-brane and
$E_N$ global symmetry comes from the 7-branes at $z=0$.
BPS states of the $E_N$ theory correspond to string webs connecting 
the 7-branes and the D3-brane.

\begin{figure}
\hspace{5cm}
\epsfbox{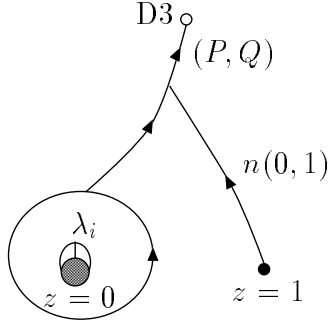}
\caption{A string web in the affine $E_N$ 7-brane background}
\label{fig:junction}
\end{figure}

BPS states of $E_N$ theory are characterized by $(N+3)$ integers
$\{P, Q, n, \lambda_i \}$ where $P$ and $Q$ are the electric 
and the magnetic charges,
$n$ represents the Kaluza--Klein modes of the $S^1$ compactification,
and $\{\lambda_i\}$ ($i=1,2,\ldots,N$) is the $E_N$ Dynkin label.
These integers are identified with the charges of string webs as 
depicted in Fig.\ref{fig:junction} where $(p,q)$ denotes an open string 
which can end on $[p,q]$ 7-branes and $(mp,mq)$ denotes $m(p,q)$-strings.
A charge vector $(P,Q)$ represents the type of a string ending on the 
D3-brane and 
$n$ is the number of strings emanating from the 7-brane at $z=1$.
The Dynkin label $\{\lambda_i\}$ is carried by strings ending on the $E_N$ 
7-branes.
The string web becomes marginally stable when the D3-brane is
located at the junction.
Thus the possible positions of the junction determine the marginal 
stability curve.

On the other hand the D3-brane--7-brane system is dual to 
IIA theory compactified on a Calabi--Yau 3-fold with a shrinking 
del Pezzo four-cycle $\bo{B}_N$\cite{HI,MOY}. 
The position of the D3-brane corresponds to the K\"ahler modulus of 
$\bo{B}_N$.
The large radius limit of $\bo{B}_N$ is taken by letting the D3-brane 
to $z=\infty$
and $\bo{B}_N$ collapses when the D3-brane is located at $z=0$.
String webs correspond to D4, D2, D0-branes wrapped on $\bo{B}_N$.
The RR charges $Q_4, Q_2, Q_0$ are related to $P, Q, n$ as follows;
\beq
Q_4=P,\,\,\,\, Q_2=Q,\,\,\,\, Q_0=n-P-\frac{1}{2}Q.
\eeq
The homology lattice of $\bo{B}_N$ contains the root lattice of the 
$E_N$ algebra, and
$\{\lambda_i\}$ is associated with D2-branes wrapped on the 2-cycles  
constructing the $E_N$ root lattice.
These duality map enables us to determine D-brane stability using 
the D3-brane probe realization.

\section{String webs in affine $E_N$ 7-brane backgrounds}

In this section we construct 3-string junctions in the affine $E_N$ 
backgrounds. 
The junction positions are determined by the BPS condition of string webs.
It is known that the BPS condition includes at least two necessity 
conditions; the self-intersection number condition and the 
geodesic condition\cite{BFy, MNS, dWZ}. 
We start with reviewing these conditions.

\subsection{BPS conditions}

The self-intersection number of a string web with the charges
$\{P,Q,n,\lambda_i\}$ is given by\footnote{
This self-intersection number is valid when the string web ends on the 
D3-brane located in the region $|z|\gg 1$.
If it is not the case, we deform the string web by moving the D3-brane to 
$|z|\gg 1$ along a contour which does not cross the string web.}
\beq
(\bo{J},\bo{J})_{(P,Q)_n}=
\frac{1}{9-N}P^2+Q^2+PQ-2nQ-\sum_{i,j=1}^{N}\lambda_i C^{ij}\lambda_j,
\label{eqn:Int00}
\eeq
where $C^{ij}$ is the inverse of the $E_N$ Cartan matrix\cite{dWZ}.
There is an another representation,
\beq
(\bo{J},\bo{J})_{(P,Q)_n}=GCD(P,Q)-2 +2g,
\label{eqn:Int0}
\eeq
where $GCD$ denotes the greatest common divisor and $g$ is the number 
of deformations which preserve the mass of the string web.
The deformation parameters have been identified with the sizes of 
string loops in the case of the flat background\cite{B,HHS,MNS}.
In case of the curved backgrounds, however, it is difficult to find the 
deformations of string webs.
Thus in this paper we assume that the existence of $g$ deformations 
and we construct a representative of string webs.
Since $g\geq 0$, the charges of a BPS string web satisfy the condition
\beq
\frac{1}{9-N}P^2+Q^2+PQ-2nQ-\sum_{i,j=1}^{N}\lambda_i C^{ij}\lambda_j
-GCD(P,Q)\geq -2.
\label{eqn:Int1}
\eeq
In the following we concentrate on string webs satisfying (\ref{eqn:Int1}).

The additional condition comes from the fact that the mass of a BPS string 
web is minimized\cite{BFy,MNS}.
For example, let us consider a $(p,q)$-string stretched along a curve 
$C$ connecting $z=z_0$ and $z=z_1$.
The mass is given by 
\beqa
M_{(p,q)}&=&\int_C\,ds\,T_{(p,q)},\CR
	 &\geq&\left|\int_{z_0}^{z_1}(p\varpi(z)-q\varpi_D(z))dz\right|,\CR
	&=& \left|pa(z_1)-qa_D(z_1)-pa(z_0)+qa_D(z_0)\right|,
\label{eqn:mass}
\eeqa
where $ds$ is (\ref{eqn:metric}),
$T_{(p,q)}=|p-q\tau(z)|/\sqrt{\mbox{Im}\tau(z)}$ is the tension of a 
$(p,q)$-string, and $a_D$ and $a$ are the Seiberg--Witten periods 
\beq
a_D(z)= \int_0^z \, dz' \varpi_D(z'),\,\,\,\,
a(z)= \int_0^z \, dz' \varpi(z').
\label{eqn:SW}
\eeq
To minimize the mass, all points on $C$ must satisfy
\beq
\mbox{Arg}\left[pa(z)-qa_D(z)-s_1\right]=\phi,
\label{eqn:BPS1}
\eeq
where $s_1=pa(z_0)-qa_D(z_0)$, and $\phi$ is a constant between $0$ and 
$2\pi$.
The argument $\phi$ determines the direction of the $(p,q)$-string 
at $z=z_0$.

Since the mass of a string web is given by the sum of the masses of all
element strings, strings constructing a BPS web must satisfy (\ref{eqn:BPS1}).
In addition, all the strings obey (\ref{eqn:BPS1}) with the same $\phi$.
To see this we consider a 3-string junction constructed from 
a $(p,q)$-string stretching between $z=0$ and $z=z_J$, $n(0,1)$-strings 
between  $z=1$ and $z=z_J$ and a $(P,Q)$-string between $z=z_J$ and 
$z=u$ where the D3-brane is located at. 
From (\ref{eqn:BPS1}) the trajectories of these strings are given by  
\beqa
\mbox{Arg}\left[pa(z)-qa_D(z)\right]&=&\phi_1,\CR
\mbox{Arg}\left[-a_D(z)-s\right]&=&\phi_2,\label{eqn:geod}\\
\mbox{Arg}\left[Pa(z)-Qa_D(z)-Pa(z_J)+Qa_D(z_J)\right]&=&\phi_3,\nonumber
\eeqa
where we have used $s\equiv -a_D(1)$ and $a(0)=a_D(0)=0$. 
Then the mass of the string web is given by
\beqa
M_{(P,Q)_n}
&=&M_{(p,q)}+|n|M_{(0,1)}+M_{(P,Q)},\CR
&=& |pa(z_J)-qa_D(z_J)|+|-na_D(z_J)-ns|\CR
&&+|Pa(u)-Qa_D(u)-Pa(z_J)+Qa_D(z_J)|,\\
&\geq& |pa(z_J)-qa_D(z_J)-na_D(z_J)-ns\CR
&&+Pa(u)-Qa_D(u)-Pa(z_J)+Qa_D(z_J)|,\CR
&=&|Pa(u)-Qa_D(u)-ns|.\nonumber
\eeqa
where we have used the charge conservation condition at $z=z_J$, 
$(P,Q)=(p,q)+n(0,1)$.
To minimize the mass $\phi_i$ ($i=1,2,3$) must satisfy 
$\phi_1=\phi_2=\phi_3$.
The result is generalized to more complicated string webs.

The condition $\phi_1=\phi_2=\phi_3$ simplifies the geodesic condition 
(\ref{eqn:geod}).
From the first and the second equations of (\ref{eqn:geod})
we find that the junction position $z=z_J$ satisfies
\beq
\mbox{Im}\frac{p a(z_J)-q a_D(z_J)}
{- a_D(z_J)-s}=0.
\label{eqn:BPS2}
\eeq
This gives the marginal stability curve on which the state $(P,Q)_n$
decays into $(p,q)_0$ and $n (0,1)_1$.
In addition the last equation of (\ref{eqn:geod}) becomes
\beq
\mbox{Arg}\left[Pa(z)-Qa_D(z)-ns\right]=\phi.
\label{eqn:BPS3}
\eeq
Similarly one can show that the element strings of a BPS string web 
satisfy (\ref{eqn:BPS3}).
We call a string satisfying (\ref{eqn:BPS3}) a $(P,Q)_n$-string throughout
this paper.
In this notation $(p,q)_0$ represents a $(p,q)$-string from $z=0$ and 
$(0,1)_1$ represents a $(0,1)$-string from $z=1$.

Recall here that $a$ and $a_D$ have a branch cut on the positive real axis 
of the $z$-plane.
In \cite{MOY} it has been shown that $\Pi(z)=(a_D, a, s)^t$ satisfies
$\Pi(z_0+i\epsilon)=M\Pi(z_0-i\epsilon)$ where $0<\epsilon\ll 1$ and  
\beq
M=\left(\ba{ccc}1&9-N&0\\-1&N-8&0\\0&0&1\ea\right),\,\,\,\,
\left(\ba{ccc}
1&9-N&0\\0&1&1\\0&0&1
\ea\right)
\label{eqn:mono0}
\eeq
for $0<z_0<1$ and $z_0>1$, respectively.
To preserve the form (\ref{eqn:BPS3}) the charge vector $(P, Q, n)^t$ 
also jumps on the branch cut to $M'(P,Q,n)^t$ where
\beq
M'=\left(\ba{ccc}
1&9-N&0\\-1&N-8&0\\0&0&1
\ea\right),\,\,\,\,
\left(\ba{ccc}
1&9-N&0\\0&1&0\\1&9-N&1
\ea\right)
\label{eqn:mono}
\eeq
for $0<z_0<1$ and $z_0>1$.

\subsection{BPS string webs and CMS} 

We can determine trajectories of strings (\ref{eqn:BPS2}) and marginal 
stability curves (\ref{eqn:BPS3}) numerically using the explicit forms 
of $a$ and $a_D$ given in \cite{MOY}. 
The results are understood qualitatively as follows.

\begin{figure}
\epsfxsize=16cm
\epsfbox{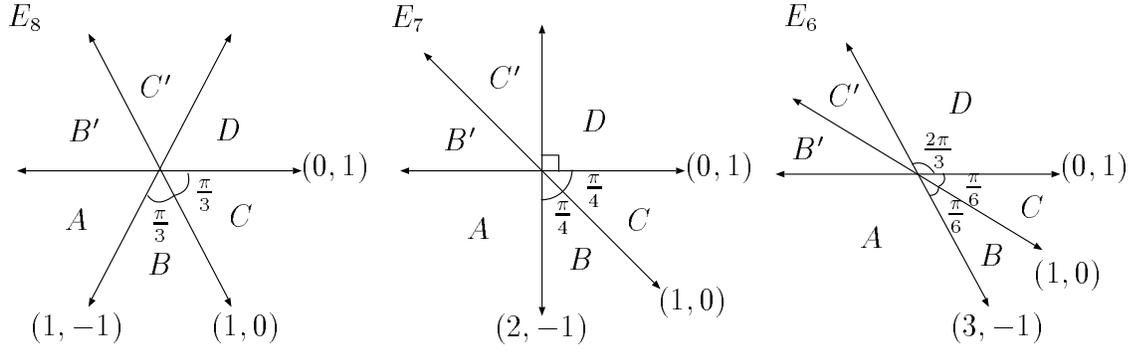}
\caption{Directions of $(P,Q)$-strings with the same $\phi$}
\label{fig:lattice}
\end{figure}

First we consider the vicinity of $z=0$. 
From (\ref{eqn:period}) we find
\beq
\left(\ba{c}\varpi_D(z)\\\varpi(z)\ea\right)\sim
\frac{\pi R^{\frac{1-\alpha}{\alpha}}\xi_2}{2(2\alpha-1)}z^{-(1-\alpha)}
\left(\ba{c}2\\-\frac{\omega}{\sin\pi\alpha}\ea\right).
\eeq
Thus $\tau=\varpi_D/\varpi$ becomes a constant $\tau_0=e^{2\pi i/3}$, 
$\sqrt{2}e^{3\pi i/4}$, $\sqrt{3}e^{5\pi i/6}$
for $E_8$, $E_7$, $E_6$, and the curve (\ref{eqn:BPS3}) becomes
\beq
\mbox{Arg}\left[(P-Q\tau_0)w-ns\right]=\phi,
\label{eqn:lines}
\eeq
where $w=c_0z^{\alpha}$ with $c_0=-\frac{\omega}{\sin\pi\alpha}
\frac{\pi R^{\frac{1-\alpha}{\alpha}}\xi_2}{2(2\alpha-1)\alpha}$.
Thus a $(P,Q)$-string stretches along a straight line on the $w$-plane 
with the angle
\beq
\mbox{Arg} [w_{(p,q)}]
=\phi-\mbox{Arg}\left[P-Q\tau_0\right].
\label{eqn:argument}
\eeq
The relative angles of various $(p,q)$-strings obeying (\ref{eqn:lines})
with the same $\phi$ are shown in Fig.\ref{fig:lattice}. 

\begin{figure}
\epsfbox{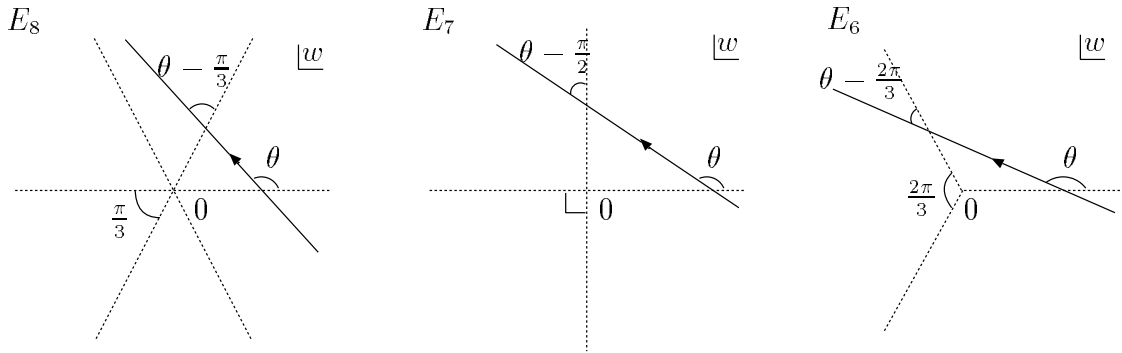}
\caption{BPS strings on the $w$-plane}
\label{fig:w}
\end{figure}

If we set the $c_0=1$, the branch cut on the $z$-plane is mapped to the
lines $\mbox{Arg}[w^{\frac{1}{\alpha}}]=0$ as shown in Fig.\ref{fig:w}.
Note that an angle between a $(P,Q)$-string and the branch cut changes 
from $\mbox{Arg}[w_{(P,Q)}]$ to $\mbox{Arg}[w_{(P,Q)}]-2\alpha\pi$
when the string crosses the branch cut(See Fig.\ref{fig:w}).
From (\ref{eqn:argument}) this means that the $(P,Q)$-string changes to 
a $(P-(9-N)Q, -P+(N-8)Q)$-string. 
This reproduces the effect of $M'$ for $0<z_0<1$ given in (\ref{eqn:mono}).

Next we introduce a $[0,1]$ 7-brane at $z=1$. 
BPS strings are not along straight lines on the 
$w$-plane, however, the order of $(P,Q)_0$-strings shown in 
Fig.\ref{fig:lattice} 
is unchanged unless the strings hit or encircle the $[0,1]$ 7-brane.
If the order of $(p_1,q_1)_0$ and $(p_2,q_2)_0$ is changed,
two strings cross at a point $z=z_0\neq 0$.
In addition if the string does not encircle the $[0,1]$ 7-brane, 
we can move the branch cut not to cross the strings.
Thus $(p_i,q_i)$ ($i=1,2$) are unchanged.
Note that a $(p_1,q_1)$-string passing through $z=z_1$ is given by 
\beq
\mbox{Arg}\left[p_1 a(z)-q_1 a_D(z)-p_1 a(z_1)+q_1 a_D(z_1)
\right]=\phi.
\label{eqn:para}
\eeq
This is parallel to $(p_1,q_1)_0$ with the same $\phi$
and it sweeps the $z$-plane as we varies $z=z_1\in i\bo{R}$.
Thus there exists a $(p_1,q_1)$-string which touches $(p_2,q_2)_0$ at 
a point $z=z_J$ between $z=0$ and $z=z_0$.
Since the direction of a $(p,q)$-string is locally given by 
\beq
\mbox{Arg}[p-q\tau(z)]=\phi,
\eeq
$z_J$ satisfies 
\beq
\mbox{Arg}[p_1-q_1\tau(z_J)]=\mbox{Arg}[p_2-q_2\tau(z_J)].
\eeq
Therefore $\tau(z_J)$ must be a rational number 
when $(p_1,q_1)$ are not parallel to $(p_2,q_2)$. 
$\tau(z_J)\in \bo{Q}$ is mapped to $i\infty$ by the $SL(2,\bo{Z})$ 
transformation of the torus thus there exists a 7-brane at $z=z_J$.
Hence the order of the strings is preserved unless the strings hit the 
7-brane. 

On the basis of the observations we will construct 3-string junctions.
3-string webs are constructed from $(p,q)_0$ and $n(0,1)_1$ 
and also $(p,q)_0$ and $n(N-9,1)_1$ which is the $n$ strings emanating
from the $[0,1]$ 7-brane at $z=1$ to the lower half $z$-plane.
From (\ref{eqn:BPS2}) we see that the trajectory of $(P,Q)_n$ with 
$\phi$ is the same with that of $(-P,-Q)_{-n}$ with $\phi+\pi$.
Thus we only consider $(P,Q)_n$-string webs with $n\geq 0$.
In what follows we will only show the analysis of the $E_8$ theory since 
the analysis is identical with that of the $E_7$ and $E_6$ theories.

\begin{figure}
\epsfbox{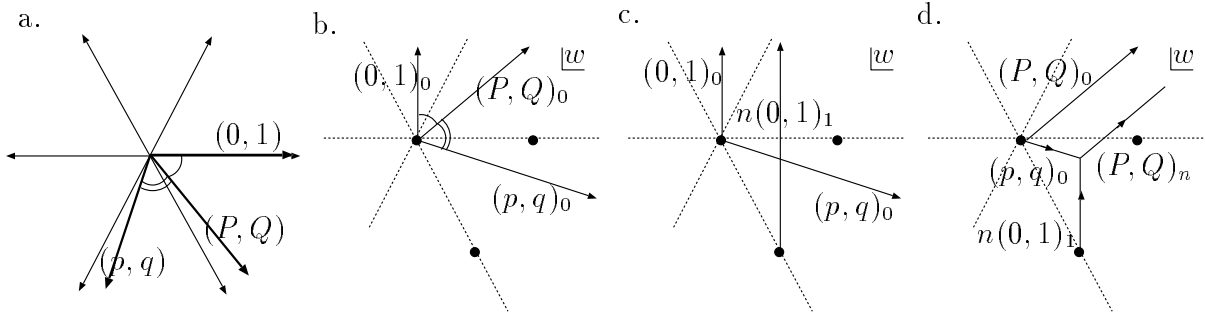}
\caption{Construction of 3-string webs}
\label{fig:wx}
\end{figure}

First we consider a $(P,Q)_n$-string web constructed from $(p,q)_0$ and
$n(0,1)_1$.
Note that the strings $(P,Q)_0$, $(p,q)_0$ and $(0,1)_0$ are approximately 
represented by straight lines from $w=0$ with the direction 
given by (\ref{eqn:argument}) as shown in Fig.\ref{fig:wx}a and b.
Since $(0,1)_1$ is parallel to $(0,1)_0$, the trajectory is qualitatively 
determined as depicted in Fig.\ref{fig:wx}c.
We see that $(0,1)_1$ can cross $(p,q)_0\in A\cup B\cup C$ in the 
Fig.\ref{fig:lattice} and can construct a junction.
Adding a line parallel to $(P,Q)_0$ to the junction we have a 
$(P,Q)_n$-string web as shown in Fig.\ref{fig:wx}d.
Note that $(P,Q)_0$ runs between $(0,1)_0$ and $(p,q)_0$ thus 
$(P,Q)\in A\cup B\cup C$.

\begin{figure}
\epsfbox{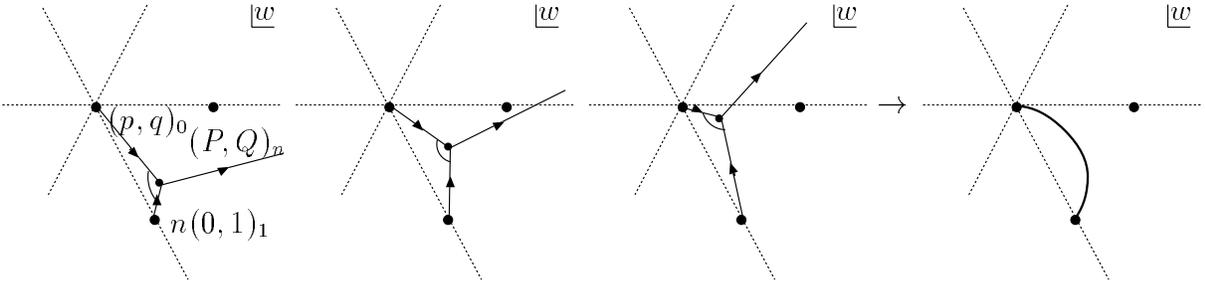}
\caption{Determination of a marginal stability curve}
\label{fig:CMSpre}
\end{figure}
\begin{figure}
\epsfbox{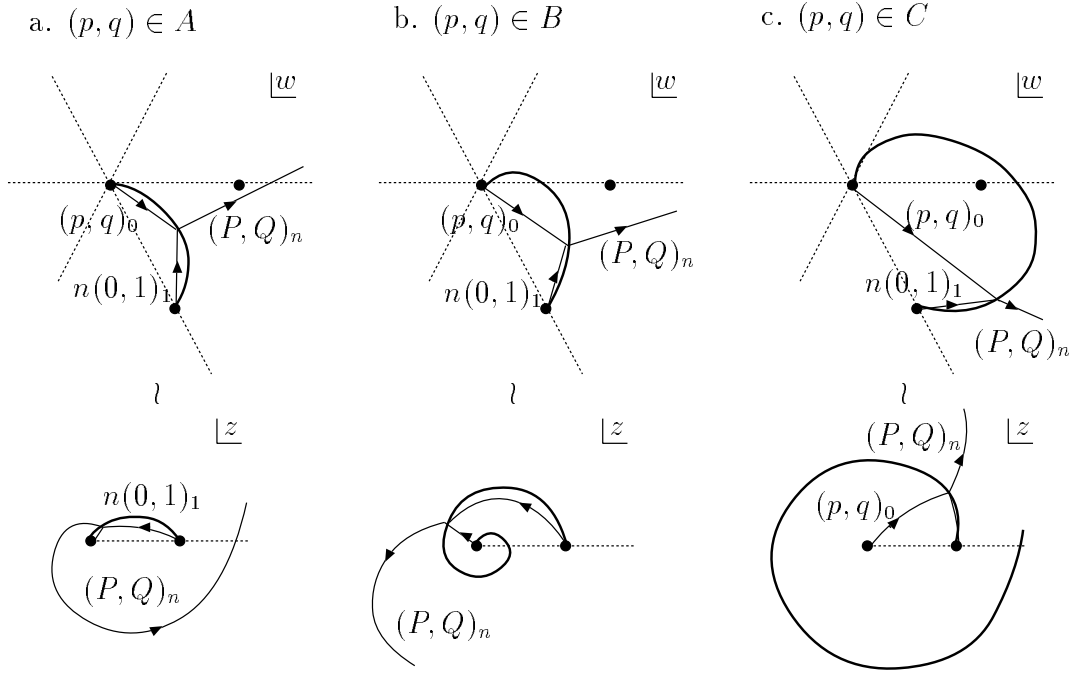}
\caption{Marginal stability curves of 3-string webs}
\label{fig:CMS}
\end{figure}

Varying the direction of $(p,q)_0$ we obtain the orbit of the junction
position (\ref{eqn:BPS2}) as depicted in Fig.\ref{fig:CMSpre}.
As a result we see that the marginal stability curves are classified 
by the charges $(p,q)$.
For a web with $(p,q)\in A$ in Fig.\ref{fig:lattice}, 
the curve is an arc from $z=1$ to $z=0$ which does not cross the branch 
cut as depicted in Fig.\ref{fig:CMS}a.
The curve for $(p,q)\in B$ is an arc from $z=1$ to $z=0$ which crosses 
the branch cut between $z=0$ and $z=1$ as shown in Fig.\ref{fig:CMS}b.
The boundary of the regions $A$ and $B$ is $(p,q)=(1,-1)$.
This is because $(1,-1)_0$ stretches along $\mbox{Arg}[w]=0$ when $(0,1)_1$ 
stretches along $\mbox{Arg}[w]=\frac{5}{3}\pi$.
This might be changed by the effect of the $[0,1]$ 7-brane, however, 
it is not the case.
Using the exact form of the periods (\ref{eqn:period}) we see that 
$(0,1)_1$ goes along $z=x+i\epsilon$ where $0<x<1$ and 
$0<\epsilon\ll 1$ when $\phi=0$.
This means that $(1,-1)$-strings with $\phi=0$
stretches along $z=x-i\epsilon$.
These strings are in agreement with the two lines in the $w$-plane.

In addition the curve for $(p,q)\in C$ starts from $z=1$, encircles
$z=0$ and crosses the branch cut in $z\geq 1$ as depicted in
Fig.\ref{fig:CMS}c.
The boundary of the region $B$ and $C$ is $(p,q)=(1,0)$.
This is because $(1,0)_0$ stretches along $\mbox{Arg}[w]=0$
when $(0,1)_1$ stretches along 
$\mbox{Arg}[w-e^{-\frac{\pi}{3}}i]=\frac{\pi}{3}$, thus $(1,0)_0$
meets $(0,1)_1$ at $z=1-i\epsilon$.
Actually we can show that the value of $\phi$ of $(1,0)_0$ passing 
through $z=1-i\epsilon$ is the same with that of $(0,1)_1$ passing 
through $z=1-i\epsilon$;
\beq
\mbox{Arg}[a(1-i\epsilon)]=
\mbox{Arg}[-a_D(1-i\epsilon)-s].
\eeq
Here we have used (\ref{eqn:mono}) and $a_D(1+i\epsilon)=-s$.

The analysis is generalized to the string webs constructed from 
$(p,q)_0$ and $n(-1,1)_1$.
In this case $(p,q)_0\in A, B', C'$ can cross $(-1,1)_n$ and construct
3-string junctions.
We can see that the trajectory of a $(P,Q)_n$-string web 
is the mirror image of that of a $(-P-Q,Q)_n$-string web constructed 
from $(-p-q,q)_0$ and $n$ $(0,1)_1$ with respect to the real axis in 
the $z$-plane.

\begin{figure}
\epsfxsize=16cm
\epsfbox{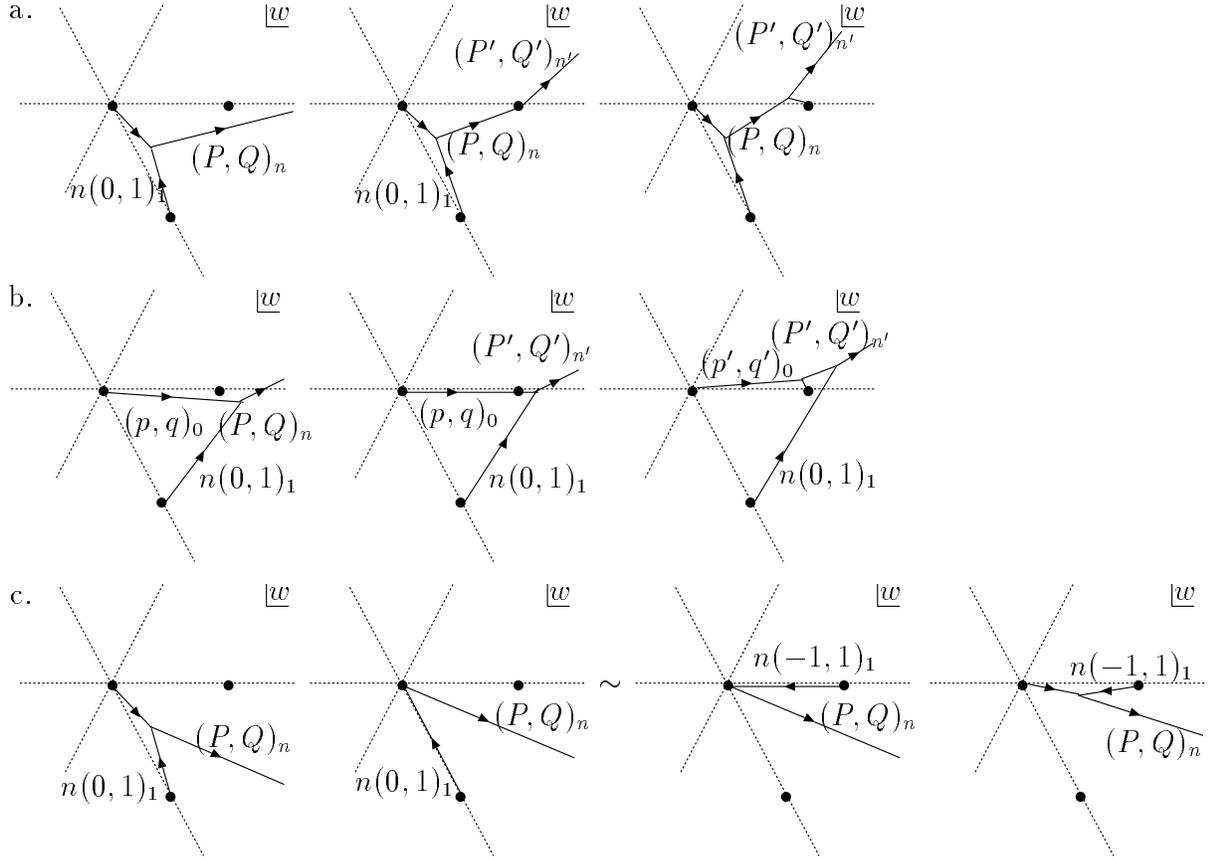}
\caption{Transitions of string webs}
\label{fig:transit}
\end{figure}

Now we determine transitions of string webs.
First let us consider a $(P,Q)_n$-string web constructed from $n(0,1)_1$ 
and $(p,q)_0$ with $(p,q)\in A \cup B$ and 
$(P,Q)=(p,q+n)\in\hspace{-3.4mm}\backslash A$.
For a certain value of $\phi$ the $(P,Q)_n$-string of the web crosses 
the branch cut in $|z|>1$ and changes to 
$(P',Q')_{n'}=(P+Q,Q)_{P+Q+n}$.
As we increase $\phi$, the string hits the $[0,1]$ 7-brane at $z=1$
and the Hanany--Witten effect creates $(P+Q)$ $(0,1)_1$-strings
as depicted in Fig.\ref{fig:transit}a\cite{GZ}.
Thus the marginal stability curve of $(P',Q')_{n'}$ is 
connected to that of $(P,Q)_n$.
The curve of $(P',Q')_{n'}$ is determined from the junction of
$(0,1)_1$ and $(P',Q')_{n'}$;
\beq
\mbox{Im}\left[\frac{P'a(z_J)-Q'a_D(z_J)-ns}{-a_D(z_J)-s}\right]=0.
\eeq
This is equal to (\ref{eqn:BPS2}) with $(p,q)=(P',Q')-n'(0,1)$,
that is, the marginal stability curve of a 3-string junction 
constructed from $n'(0,1)_1$ and $(P',Q'-n')_0$.

Second we consider a $(P,Q)_n$-string web constructed from $n(0,1)_1$ and
$(p,q)\in C$.
At a certain value of $\phi$ $(P,Q)_n$, $(p,q)_0$ and $(0,1)_1$ cross the 
branch cut in $|z|>1$ and change to  $(P',Q')_{n'}=(P+Q,Q)_{P+Q+n}$, 
$(p',q')_{l}=(p+q,q)_{p+q}$ and $(1,1)_2$.
As we increases $\phi$ $(p,q)_0$ hits the 7-brane at $z=1$ and 
the Hanany--Witten effect creates 
$(0,1)_1$-strings as depicted in Fig.\ref{fig:transit}b.
Thus the marginal stability curve of $(P',Q')_n'$ is given by 
\beq
\mbox{Im}\left[\frac{P'a(z_J)-Q'a_D(z_J)-n's}{p'a(z_J)-q'a_D(z_J)-ls}\right]
=0.
\eeq
This curve is connected to the marginal stability curve for $(P,Q)_n$
depicted in Fig.\ref{fig:CMS}c.

Finally we consider a $(P,Q)_n$-string web with $(P,Q)\in A$.
When the string web is constructed from $n(0,1)_1$ and $(p,q)_0$,
the $(P,Q)_n$-string of the web does not cross the branch cut as shown 
in Fig.\ref{fig:transit}c.
On the other hand the strings hit the 7-branes at $z=0$ at
a certain value of $\phi$.
As we increase $\phi$ the web changes to the $(P,Q)_n$-string web 
constructed from $n(-1,1)_1$ and $(P+n,Q-n)_0\in A$.
The marginal stability curve is the mirror image of Fig.\ref{fig:CMS}a.
Thus the marginal stability curve of $(P,Q)_n$ becomes a circle passing 
through $z=0$ and $z=1$.

These transitions give us a set of marginal stability curves which are 
connected 
as was shown in four-dimensional $SU(2)$ theories\cite{BFy2,O}. 
In this case, however, there appear two unusual behaviors.
First, if $(P,Q)\in B\cup C$, the $(P,Q)_n$-string of a web constructed 
from $n(0,1)_1$ and $(p,q)_0\in A\cup B$ passes through the branch cut 
on $0<z<1$ and intersects $(0,1)_1$ of the web. 
Are such unusual string webs actually BPS?
We will approach the question in the next section.
Second, two or more marginal stability curves would appear for a state.
For example, a $(3,2)_4$-string web constructed from
$4(0,1)_1$ and $(3,-2)_0$ has a marginal stability curve
depicted in Fig.\ref{fig:CMS}b.
On the other hand a marginal stability curve of $(1,2)_1$-string web 
constructed from $(0,1)_1$ and $(1,1)_0\in C$ is connected to that 
of $(3,2)_4$.
The curve starts from $z>1$ thus it is different from the curve 
depicted in Fig.\ref{fig:CMS}b.
We will resolve the multiplicity in section 5 by generalizing the 
self-intersection number condition.

\section{Open string -- string loop transition}
\begin{figure}
\hspace{3.4cm}
\epsfbox{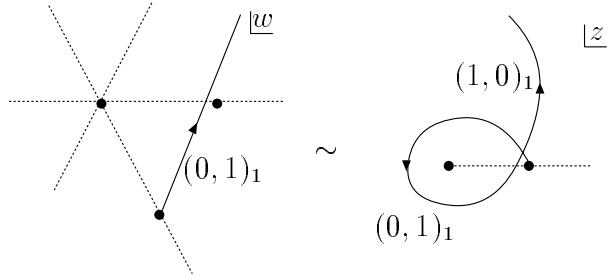}
\caption{An open string intersecting itself}
\label{fig:01}
\end{figure}

In the previous section we have seen that some string webs intersect 
themselves.
The simplest example is a $(0,1)_1$ depicted in Fig.\ref{fig:01} for 
the $E_8$ case.
In this section we will show that the string web is BPS.

\begin{figure}
\hspace{1.4cm}
\epsfbox{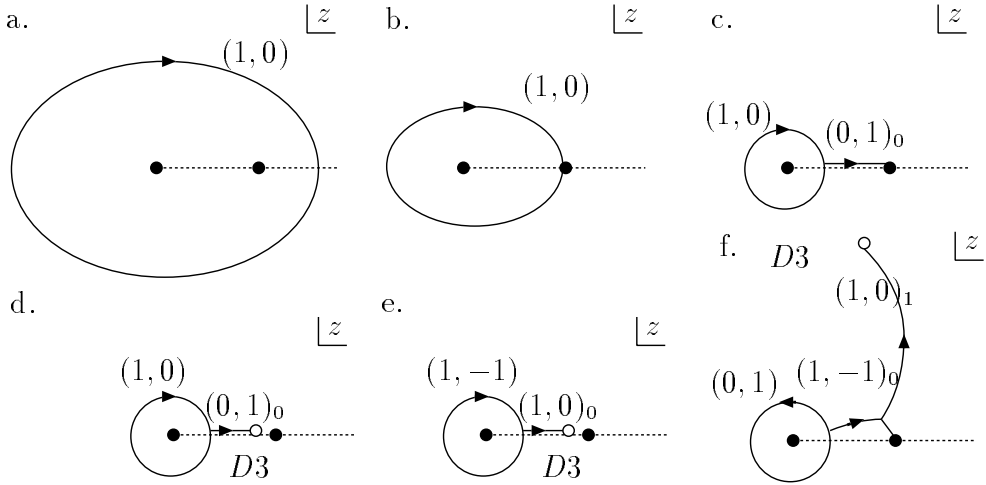}
\caption{Relation between the loop string and the $(1,0)_1$-string web}
\label{fig:loop}
\end{figure}

For this purpose recall that the affine $E_8$ 7-brane system is T-dual 
to an O8-plane and seven D8-branes compactified on $S^1$.
In the 8-brane background there exists a string winding on $S^1$.
The string is dual to a $(1,0)$-string drawing a loop around the 7-branes.
Thus the loop string is BPS at least in the semi-classical region 
$|z|\gg 1$.
The trajectory of the loop string crossing the positive real axis 
of the $z$-plane at $z=z_0>1$ is given by 
\beq
\mbox{Arg}[-a(z)+a(z_0+i\epsilon)]
=\mbox{Arg}[-a(z_0-i\epsilon)+a(z_0+i\epsilon)]=\mbox{Arg}[s],
\eeq
where $0<\epsilon\ll 1$.
In the second equality we have used (\ref{eqn:mono0}). 
The explicit form of the curve is determined by numerical calculations. 
The result is a loop as shown in Fig.\ref{fig:loop}a.
As we decrease the value of $z_0$, the size of the loop string becomes
smaller.
From (\ref{eqn:mass}) and (\ref{eqn:mono0}) we obtain the mass of the 
loop string, 
\beq
M_{\mbox{{\scriptsize loop}}}
=|a(z_0+i\epsilon)-a(z_0-i\epsilon)|=|s|.
\eeq
The mass is independent of $z_0$.
This reflects the fact that the mass of the string winding on $S^1$ is
independent of the distance from 8-branes.

As we decreases $z_0$ further, the loop string hits the 7-brane at $z=1$.
Then Hanany--Witten effect creates a $(0,1)$-string ending on $z=1$,
and the loop string becomes the string web as depicted in 
Fig.\ref{fig:loop}c.
When the loop around $z=0$ crosses the branch cut $z=z_0<1$, 
the mass is given by 
\beqa
M_{\mbox{{\scriptsize loop}}}
&=&|a(z_0+i\epsilon)-a(z_0-i\epsilon)|+|a_D(z_0+i\epsilon)+s|,\CR
&\leq&|a(z_0+i\epsilon)-a(z_0-i\epsilon)+a_D(z_0+i\epsilon)+s|,\CR
&=&|s|.\nonumber
\eeqa
Thus the mass is preserved under the creation of the $(0,1)_1$-string
and independent of the loop size.

Now we introduce a D3-brane probe on the $(0,1)$-string.
Then we find that a $(0,1)_0$-string depicted in Fig.\ref{fig:loop}d exists.
The string web involves a loop and the mass of the web is independent of
the loop size.
Next we move the D3-brane anti-clockwisely and make it cross the 
branch cut in $|z|<1$.
The $(0,1)_0$ undergoes the effect of $M'$ given in (\ref{eqn:mono})
and becomes $(1,0)_0$ involving a loop as depicted in Fig.\ref{fig:loop}e.
Finally we move the D3-brane further and make it cross the branch cut 
in $|z|>1$.
The $(1,0)_0$-string undergoes the effect of $M'$ and becomes a 
$(1,0)_1$-string.
When the $(1,0)_1$-string hits the 7-brane at $z=1$,
a $(0,1)_1$-string is created by the Hanany--Witten effect as depicted in 
Fig.\ref{fig:loop}f.
We find that the string depicted in Fig.\ref{fig:01} appears 
when the loop size is maximal.
Thus the string web shown in Fig.\ref{fig:01} is BPS.

Note that the string web has a deformation parameter corresponding to the 
size of the loop.
Such parameters have been related to the zero-modes of a BPS state in the 
field theory\cite{N=4}.
As we move the D3-brane along the $(1,0)_1$-string web in Fig.\ref{fig:01}
and locate the D3-brane at the vertex of two string, the web becomes 
marginally stable.
The marginal stability curve is given by (\ref{eqn:BPS2})
with $(p,q)=(1,-1)$.
If the size of the loop is not maximal, the string web disappears 
from the BPS spectrum inside the curve.
On the other hand the string web at the maximal loop size can change to
an open string emanating from $z=1$ inside the curve\footnote{
Inside the marginal stability curve the masses of states $(1,0)_1$, 
$(1,-1)_0$ and $(0,1)_1$ satisfy the inequality $M_{(1,0)_1}> 
M_{(1,-1)_0}+M_{(0,1)_1}$. 
The $(1,0)_1$ state, however, still has a BPS representative of string 
webs.}.
In this transition the string web loses the deformation parameter 
thus the transition would be detected as the annihilation of zero-modes 
of the BPS state.

The number of deformation parameters $g$ is determined by 
(\ref{eqn:Int00}) and (\ref{eqn:Int0}).
The results reproduce the analysis above.
In addition $g$ of an arbitrary self-intersecting string web 
decreases when the string web loses the self-intersection.
For example $g$ of a $(P,Q)_n$ self-intersecting string web involving
$(0,1)_1$-strings is given by 
\beq
2g=(\bo{J},\bo{J})_{(P,Q)_n}-GCD(P,Q)+2.
\eeq
When the D3-brane moves across the branch cut in $|z|<1$ from the 
upper-half $z$-plane, the self-intersection of the web disappears.
Simultaneously the number of the deformation parameter $g'$ becomes
\beq
2g'=(\bo{J},\bo{J})_{(-Q,P+Q)_n}-GCD(-Q,P+Q)+2=2g-2nP.
\eeq
Since $2nP<0$, we see that the self-intersection number decreases.
Thus the interpretation of the transition would be 
understood as that of the $(1,0)_1$-string web.

\section{Local self-intersection number condition and relevant CMS}

We have noted in section 3 that the existence of all the 3-string 
junctions causes multiplicity of marginal stability curves.
The multiplicity is resolved if we apply the self-intersection number
condition on all parts of a string web.
The reason why the additional condition is needed can be seen by 
introducing additional D3-brane probes.
When the D3-branes are located on a string web, the string web is 
divided into many parts. 
Each of them can freely moves along the world-volume of the D3-branes.
Thus any parts of the string web must be BPS and satisfy the 
self-intersection number condition.

A 3-string junction constructed from $n(0,1)_1$ and $(p,q)_0$ is 
divided into two parts by introducing an extra D3-brane on $(p,q)_0$.
The self-intersection number condition for $(p,q)_0$ is satisfied
if the web satisfies (\ref{eqn:Int1}).
The self-intersection number of the other part is given in \cite{Iqbal}
as $(\bo{J},\bo{J})=n(P-n)$, and the self-intersection number condition 
becomes
\beq
n(P-n)\geq -2+ GCD(P,Q) + GCD(p,q),
\eeq
where $(P,Q)_n=(p,q)_0+n(0,1)_1$.
One can prove that the condition is equivalent to
\beq
n(P-n)\geq 0,
\label{eqn:Int2}
\eeq
for the string webs with $GCD(P,Q,n)=1$.
For the webs with $n>0$ (\ref{eqn:Int2}) is rewritten as $0<n\leq P$.
Other string webs with the charges $P>n$ are 
constructed by deforming the 3-string webs with $0<n\leq P$
as we have done in the last part of section 3.
One can show that a $(P,Q)_n$-string web is uniquely determined
in this way if $(P_l,Q_l)\in\hspace{-3.4mm}\backslash
D$ for $\forall l\in \bo{Z}$ where $(P_l,Q_l,n_l)^t=M'^{-l}(P,Q,n)$
with $M'$ in (\ref{eqn:mono}) for $|z_0|>1$.

\begin{figure}
\hspace{1.4cm}
\epsfbox{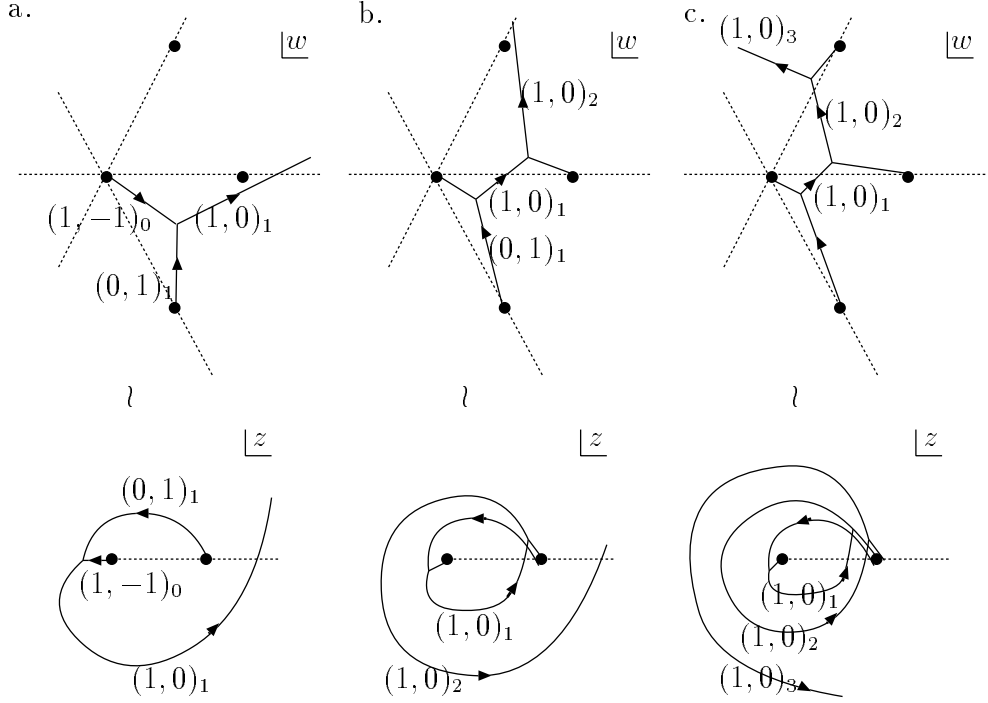}
\caption{$(1,0)_n$-string webs} 
\label{fig:KK}
\end{figure}

For example we explicitly construct a $(1,0)_3$-string web in 
the $E_8$ theory.
When the web is in the fundamental representation of $E_8$, the web 
corresponds to a Kaluza-Klein mode of a quark in the five-dimensional 
theory. 
We see that $(P,Q,n)^t=M'^{-l}(1,0,3)^t$ satisfies the condition 
(\ref{eqn:Int2}) when $l=2$ and $(P,Q,n)=(1,0,1)$.
Thus the $(1,0)_3$ is constructed by deforming $(1,0)_1$.
The $(1,0)_1$-string web is a 3-string junction constructed from 
$(0,1)_1$ and $(1,-1)_0$ as depicted in Fig.\ref{fig:KK}a. 
The marginal stability curve is as shown in Fig.\ref{fig:CMS}a.
As we increases $\phi$ the $(1,0)_1$-string crosses the branch cut 
in $|z|>1$ and the $(1,0)_1$-string becomes a $(1,0)_2$-string.
When the string hits the 7-brane at $z=1$ the Hanany--Witten effect 
creates a $(0,1)_1$-string as depicted in Fig.\ref{fig:KK}b.
The $(1,0)_2$-string web decays into $(0,1)_1$ and $(1,-1)_1$
when the D3-brane is located at the junction.
The marginal stability curve is given by (\ref{eqn:BPS2}) with 
$(p,q)=(1,-2)$ and determined as shown in Fig.\ref{fig:CMS}a.
As $\phi$ increases further, we can construct the $(1,0)_3$-string web
as shown in Fig.\ref{fig:KK}c.
The marginal stability curve is given by (\ref{eqn:BPS2}) with 
$(p,q)=(1,-3)$.
Similarly we can construct a $(1,0)_n$-string web involving $(n+1)$
junctions.
The length of the $(0,1)_1$-string created in the $n$-th 
Hanany--Witten effect is shorter than that of $(n-1)$.
Thus the marginal stability curve of the state $(1,0)_{n+1}$ is nearer
to the real axis of the $z$-plane than that of the state $(1,0)_{n}$.

Mapped the string webs to the $z$-plane we find that string webs involve
inner-surfaces which look like loops(See Fig\ref{fig:KK}).
The sizes of the inner surfaces, however, are fixed.
To enable to change the size of inner surfaces, all strings must be 
merged at junctions, but this breaks the self-intersection number
condition.
The same phenomenon has been already known in the D3-brane probe 
realization of four-dimensional $SU(2)$ QCD\cite{O}.

Similarly we can find the condition $0<n\leq -(P+Q)$
of a $(P,Q)_n$ 3-string web constructed from $n(-1,1)_1$ and $(p,q)_0$.
The explicit forms of string webs generated from the 3-string webs 
are determined by the mirror image of that of a $(-P-Q,Q)_n$-string web 
involving $(0,1)_1$-strings with respect to the real axis in 
the $z$-plane.

\section*{Acknowledgements}
I would like to thank S.~K.~Yang for helpful discussions. This work was 
supported by JSPS Research Fellowships for Young Scientists.


\end{document}